\documentclass[usenatbib]{mn2e}

\usepackage{graphicx}
\usepackage{psfig}
\usepackage{epsfig}
\usepackage{natbib}

\newcommand\xmmsrc{2XMM~J202131.0+402645}
\newcommand\psr{PSR~J2021+4026}

\title[Study of the $\gamma$-ray pulsar \psr]{Multiwavelength properties of a new Geminga-like pulsar: \psr}
\author[L. Trepl et al.]{L. Trepl$^{1}$\thanks{E-mail: 
ltrepl@astro.uni-jena.de}, C.Y. Hui$^{2,3}$, 
K.S. Cheng$^{3}$, J. Takata$^{3}$, Y. Wang$^{3}$, Z.Y. Liu$^{4}$ and N. Wang$^{4}$\\
$^{1}$Astrophysikalisches Institut und Universit\"ats-Sternwarte, Universit\"at Jena,
Schillerg\"a\ss chen 2-3, 07745 Jena, Germany\\
$^{2}$Department of Astronomy and Space Science, Chungnam National University,
Daejeon, South Korea\\
$^{3}$Department of Physics, University of Hong Kong, Pokfulam Road, Hong
Kong\\
$^{4}$Urumqi Astronomical Observatory, NAO-CAS, 40 South Beijing Road, Urumqi, 830011, China\\}

\begin{document}

\date{Accepted 2010 February 15. Received 2010 February 11; in original form 2009 November 19}

\pagerange{\pageref{firstpage}--\pageref{lastpage}} \pubyear{2010}

\maketitle

\label{firstpage}

\begin{abstract}
In this paper, we report a detailed investigation of the multiwavelength properties 
of a newly detected $\gamma-$ray pulsar, \psr, in both observational and theoretical aspects. 
We firstly identify an X-ray source in the \emph{XMM-Newton\/} serendipitous source catalogue, \xmmsrc, 
located within the $95\%$ confidence circle of \psr. 
With an archival \emph{Chandra} observation, 
this identification provides an X-ray position with arcsecond accuracy 
which is helpful in facilitating further investigations. 
Searching for the pulsed radio emission at the position of \xmmsrc\ 
with a 25-m telescope at Urumqi Astronomical Observatory resulted in null detection and places 
an upper-limit of 0.1~mJy for any pulsed signal at 18~cm. 
Together with the emission properties in X-ray and $\gamma-$ray, the radio quietness suggests \psr\ 
to be another member of Geminga-like pulsars. In the radio sky survey data, 
extended emission features have been identified in the $\gamma-$ray error circle of \psr. 
We have also re-analyzed the $\gamma-$ray data collected by \emph{FERMI}'s Large Area Telescope. 
We found that the X-ray position of \xmmsrc\ is consistent with that of the optimal $\gamma-ray$ timing solution.  
We have further modeled the results in the context of outer gap model which provides us with constraints for the pulsar 
emission geometry such as magnetic inclination angle and the viewing angle. We have 
also discussed the possibility of whether \psr\ has any physical association with the supernova 
remnant G78.2+2.1 ($\gamma-$Cygni).  
\end{abstract}


\section{Introduction}

The successful launch of the \emph{FERMI} Gamma-Ray Space Telescope has led us into a
new era of high-energy astrophysics. The sensitivity of the Large Area Telescope (LAT) on the spacecraft
is much higher than that of its predecessor, the Energetic Gamma-Ray Experiment Telescope
on the \emph{Compton\/} Gamma-Ray Observatory. This enables very efficient
searches for $\gamma-$ray pulsars. Shortly after its launch, it has already expanded 
the population of $\gamma-$ray pulsars considerably (Abdo et al. 2009a,b,c). 
Abdo et al. (2009a) has further reported the detections of 16 pulsars with high significance in a
blind search, including \psr.

The $\gamma-$ray detection of \psr\ was firstly reported in the \emph{FERMI} bright source list with a 
signal-to-noise ratio $>10\sigma$ (Abdo et al. 2009d). The nominal $\gamma-$ray position of \psr\ is located 
at the edge of the supernova remnant G78.2+2.1 (Abdo et al. 2009d; Green 2009). 
Using the first 6 months of the LAT data, the timing 
ephemerides of the pulsar were recently reported by Abdo et al. (2009c). It has a spin period of $P=265$~ms and 
a spin-down rate of $\dot{P}=5.48\times10^{-14}$~s~s$^{-1}$. These spin parameters imply a characteristic age of 
$\tau_{\rm sd}\sim77$~kyr, a surface magnetic field of $\sim4\times10^{12}$~G and a spin-down luminosity 
of $\dot{E}\sim10^{35}$~erg~s$^{-1}$. 

Apart from studying the pulsar properties in the $\gamma-$ray regime, the effort in searching for counterparts 
in other wavelengths is also very important. The broadband emission properties of pulsars,  
from radio to $\gamma-$ray, are crucial in discriminating different competing models 
(e.g. see the recent reviews by Cheng 2009; Harding 2009).
This will definitely shed detailed light on many unsettled debates, such as whether the high-energy emission and radio 
emission are originated from the same accelerating region.

In this paper, we report the results from the investigation of the multiwavelength properties of \psr. In Section 2, 
we describe the searches for the possible X-ray and radio counterpart for the pulsar with both archival and dedicated 
observations. We have also analyzed all the first-year $\gamma-$ray data collected by LAT in order to constrain its 
spectral and temporal properties. In Section 3, we model the emission properties of \psr\ in the context of the outer gap model. 
We have further discussed if it is possible that \psr\ has any association with G78.2+2.1. 
Finally, we summarize our results in Section 4. 

\section[]{Data Analysis}

\subsection{X-ray analysis of the point sources in the error circle of \psr}

In order to search for possible X-ray counterparts of \psr\ 
detected in a blind search with \emph{FERMI\/} LAT,
we firstly cross-correlated the second \emph{XMM-Newton\/} serendipitous source catalogue (hereafter XMM~SSC)
(Watson et al. 2009) with the \emph{FERMI\/} LAT bright $\gamma$-ray source list, in which \psr\ 
is denoted as 0FGL~J2021.5+4026 (Abdo et al. 2009a).
XMM~SSC is the largest catalogue ever constructed in X-ray astronomy, which contains 221012 unique sources
\footnote{The incremental version of this catalogue is used throughout in this study which is available at
http://xmmssc-www.star.le.ac.uk/Catalogue/xcat\_public\_2XMMi.html}.
In particular, we search for all the X-ray sources in XMM~SSC which are
located within its $95\%$ confidence circle. 

\begin{figure}
\centering
  \includegraphics[width=9.2cm]{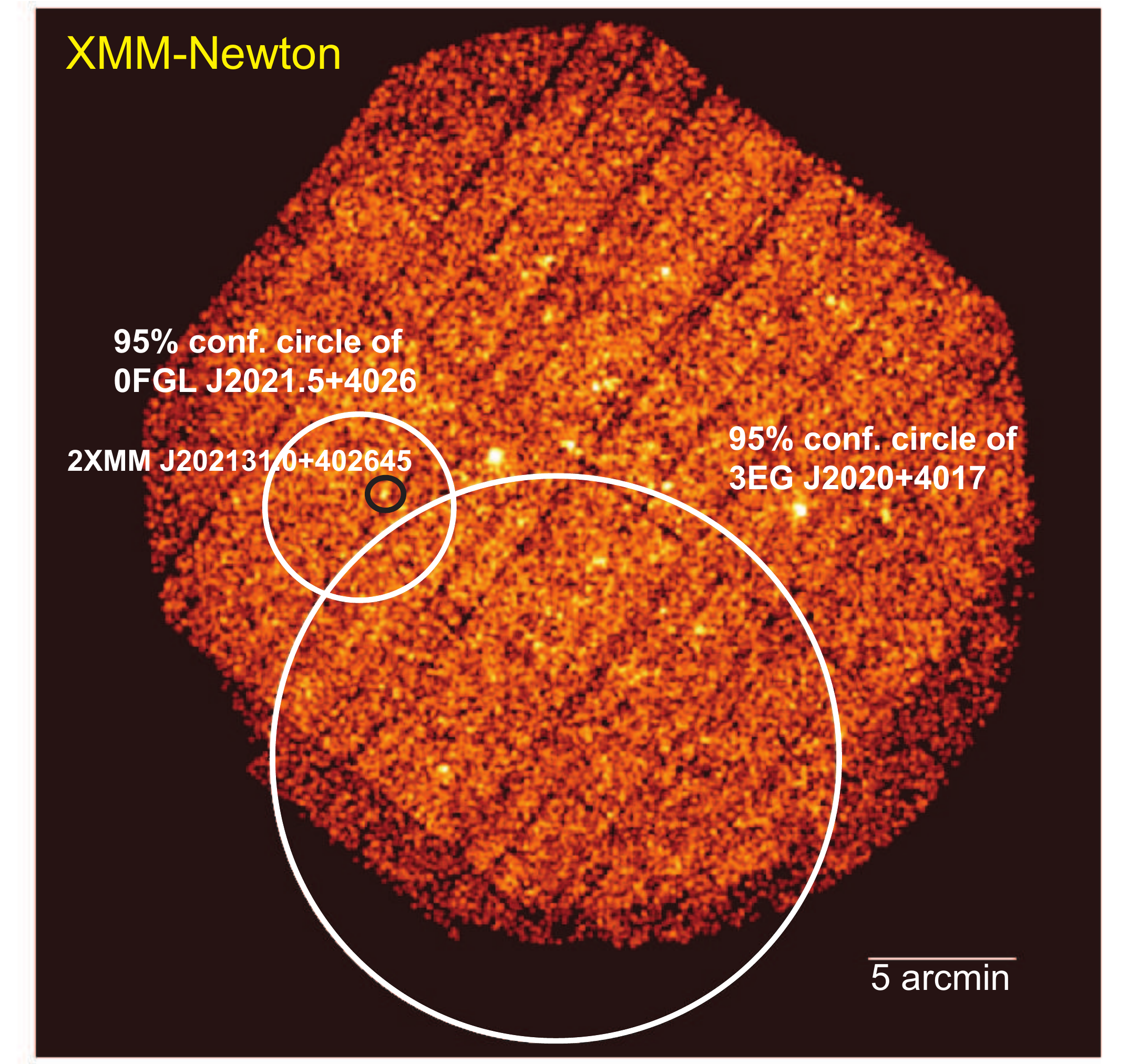}
  \caption[]{The image as observed by \emph{XMM-Newton\/} in the energy range $0.3-10$~keV 
  on 1 December 2003 with the MOS1/2 and PN data merged. The 
  aim-point of this observation was towards the geometrical center of SNR~G78.2+2.1. 
  The $95\%$ confidence circles of the $\gamma-$ray sources \psr\ and 3EG~J2020+4017 are illustrated as white circles.
  The small black circle indicates the position of \xmmsrc.}
  \label{lat_ssc}
\end{figure}

\begin{figure}
\centering
 \includegraphics[width=9.2cm]{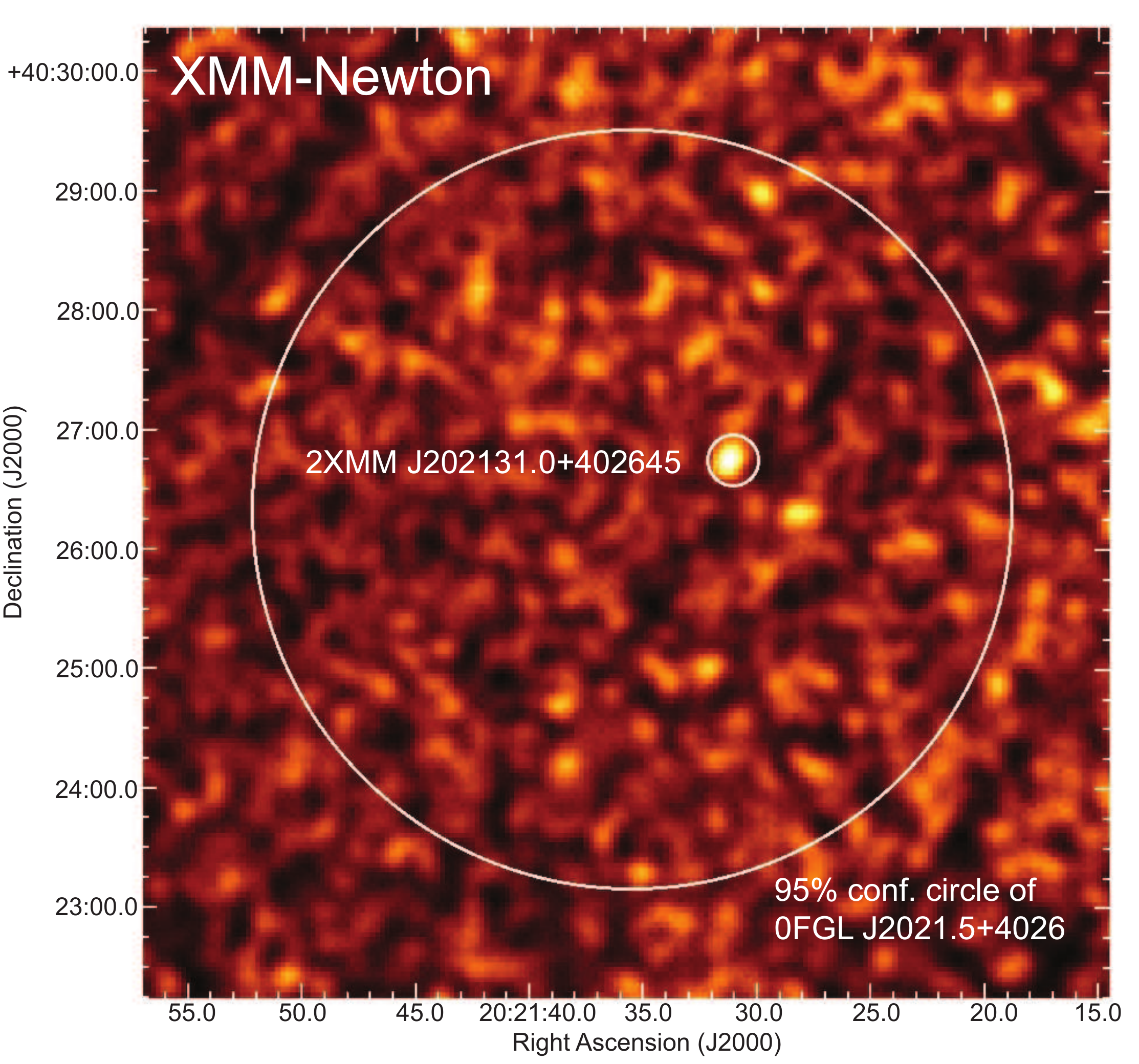}
\caption[]{The $8\times8$ arcmin$^{2}$ \emph{XMM-Newton\/} image in the energy range $0.3-10$~keV 
centered at the nominal $\gamma-$ray position of \psr.
\xmmsrc\ is the only source detected in the 95\% confidence circle of the pulsar.}
  \label{xmm_closeup}
\end{figure}

In this search, we have identified only one X-ray object, \xmmsrc, in the $95\%$ error circle of \psr\ 
with a radius of $0.053^{\circ}$. 
This X-ray source is located less than 1 arcmin away from the reported $\gamma-$ray position of the
LAT pulsar. The corresponding \emph{XMM-Newton\/} observation was carried out on 1 December 2003 with
MOS1/2 (Metal Oxide Semicondutor) and EPIC-PN (the European Photon Imaging Camera-Positive Negative) detectors 
operated in full frame mode (Obs.~ID: 150960801). This observation was
pointed to the geometrical center of the supernova remnant G78.2+2.1 with \xmmsrc\ located $\sim8.5$ arcmin off-axis.
Examining this dataset for times of high background, we notice that this observation was contaminated by
soft-proton flares. Cleaning the data by removing these flares results in the effective exposure of $\sim5$~ks
and $\sim2.6$~ks in MOS1/2 and PN detectors respectively.
The combined MOS1/2+PN image of the full field-of-view is
displayed in Figure~\ref{lat_ssc} with the error circles of \psr\ and 3EG~J2020+4017 illustrated. 
An $8\times8$~arcmin$^{2}$ close-up view
centered at the nominal position of \psr\ is displayed in Figure~\ref{xmm_closeup}.
\xmmsrc\ is located just outside the $95\%$ confidence circle of 3EG~J2020+4017 (see Fig.~\ref{lat_ssc}) which is the
brightest unidentified $\gamma$-ray source discovered by \emph{EGRET\/} (Hartman et al. 1999).

\begin{figure}
\centering
 \includegraphics[width=9.2cm]{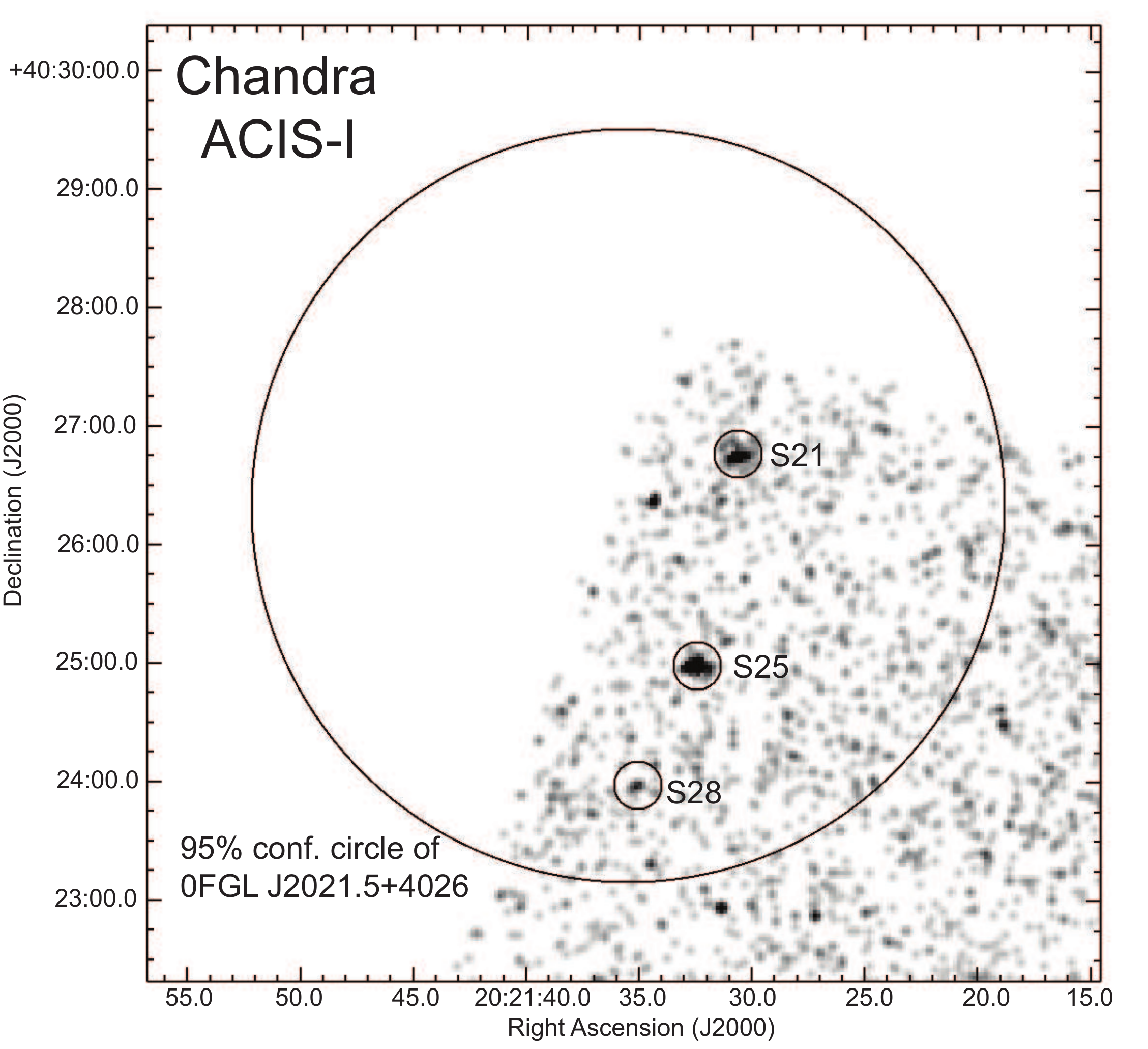}
\caption[]{The $8\times8$ arcmin$^{2}$ field centered at the nominal $\gamma-$ray position of \psr\ which is partly covered
by a \emph{Chandra} observation. The displayed image is in the energy range $0.5-8$~keV. 
Three sources located in the 95\% confidence circle of \psr\ are labeled with
the same designations in Weisskopf et al. (2006).}
\label{cxc_closeup}
\end{figure}

Searches for X-ray counterparts of 3EG~J2020+4017 have been reported by Becker et al. (2004) and Weisskopf et al. (2006)
with \emph{Chandra\/} Advanced CCD Imaging Spectrometer (ACIS) spectro-imaging observations. 
The observation made by Weisskopf et al. (2006) partly covered the 
error circle of \psr. This \emph{Chandra\/} ACIS-I observation took place on 6 February 2005 with an effective exposure
time of $\sim14$ ksec (Obs. ID: 5533). The ACIS-I image centered at the nominal $\gamma-$ray position of \psr\ is displayed
in Figure~\ref{cxc_closeup}, which has the same field-of-view as that in Figure~\ref{xmm_closeup}.
The position of source S21 (as labeled in Weisskopf et al. 2006) is found to be consistent with
that of \xmmsrc, and hence we concluded that they are the same object.
As it is located outside the error circle of 3EG~J2020+4017, it did not receive enough attention in Weisskopf et al. (2006). 

We note that two other X-ray sources detected by Weisskopf et al. (2006), namely S25 and S28 in their paper,
are also found to be located within the $95\%$ error circle of \psr\ (see Fig.~\ref{cxc_closeup}).
As a number of important parameters in characterizing these sources,
such as the estimate of the source extent and the spectral properties, were not reported by Weisskopf et al. (2006),
we decided to re-analyze this dataset.
We have independently performed the source detection with the \emph{Chandra} data by means of a wavelet algorithm.
To constrain their source positions accurately, aspect offset,
which is a function of the spacecraft roll angle, has been carefully checked and corrected prior to the analysis.
The source detection reports the signal-to-noise ratios of S21, S25 and S28 to be $5.8\sigma$, $9.6\sigma$ and
$2.5\sigma$ respectively. All three sources are found to have their spatial extents comparable with the estimated
size of the point spread function (PSF) at their off-axis angles in the CCD. Therefore, based on this \emph{Chandra}
observation, no evidence for any extended structure is found to associate with these sources.

\begin{figure}
\centering
\includegraphics[width=9.2cm]{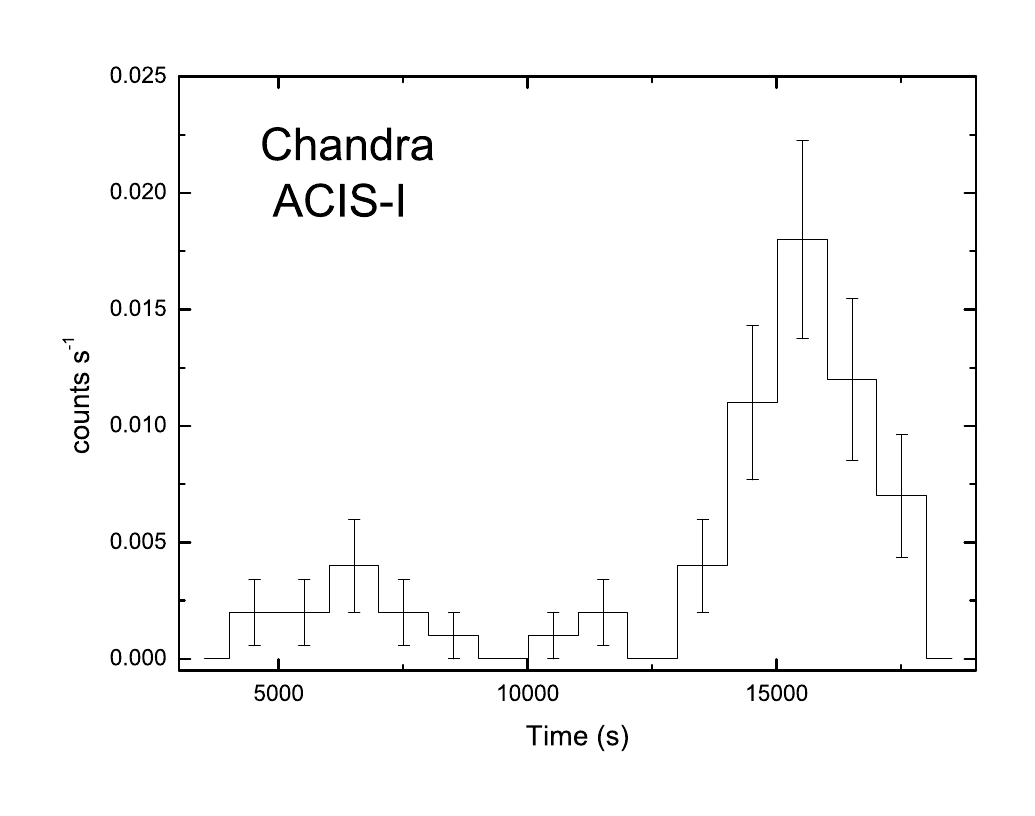}
\vspace{-0.5cm}
\caption[]{The light curve of the X-ray source S25 detected by Chandra ACIS-I which has shown its variability during 
the observation.}
\label{s25_lc}
\end{figure}

We noted that the X-ray emission from S25 is found to be highly variable in the exposure of \emph{Chandra} 
(see Figure~\ref{s25_lc}). 
Its long-term variability is further supported by the non-detection of any source at its position in the \emph{XMM-Newton}
observation.  S28 is a faint X-ray source close to the detection threshold and found to have an optical
counterpart (cf. Weisskopf et al. 2006).
These properties make these two sources very unlikely have any association with a neutron star/rotation-powered pulsar
(see the discussion in Hui \& Becker 2009).
On the other hand, no optical counterpart has been identified with S21 (cf. Tab.~5 in Weisskopf et al. 2006).
Also, we do not find any convincing evidence of temporal/spectral variability for this source. Therefore, with the 
existing data, 
S21 remains to be the \emph{only} promising X-ray counterpart of \psr\ located in the $95\%$ confidence circle of the
$\gamma$-ray pulsar.

Utilizing the \emph{Chandra} data, the X-ray position of S21 is constrained to be
RA (J2000) $=20^{h}21^{m}30\fs553$, Dec (J2000) $=+40^{\circ}26'46\farcs89$. The corresponding positional error
in each direction reported by our source detection are $\delta$RA=1.11" and $\delta$Dec=0.73" respectively.
To avoid under-estimating the uncertainties, we have to further consider the nominal
pointing uncertainty of the spacecraft. The uncertainty can be estimated
from the distribution of aspect offset for a sample of point sources with accurately
known celestial positions\footnote{http://cxc.harvard.edu/cal/ASPECT/celmon/}. There is $68\%$
of 70 sources imaged on ACIS-I have offsets smaller than $\sim0.4"$. 
We adopted this value as the astrometric
uncertainty and added to the aforementioned quoted errors in quadrature for each coordinate. This gives the
resultant $1\sigma$ positional errors as $\delta$RA=1.18" and $\delta$Dec=0.84".

Although the search for the optical counterpart of S21 by Weisskopf et al. (2006) yield null-detection, we
independently looked for any optical identification of this X-ray source in the United States Naval Observatory (USNC)-B1.0 
catalogue (Monet et al. 2003) and the Digitized Sky Survey with the improved X-ray position. 
Within our estimated $3\sigma$ X-ray positional uncertainty, 
we cannot identify any optical counterpart of S21 down to the limiting
magnitude of USNO-B1.0 catalogue (i.e. 21; cf. Monet et al. 2003).
This confirms the result reported by Weisskopf et al. (2006).

To examine the X-ray emission nature of \xmmsrc\ (S21), 
we make use of both \emph{XMM-Newton\/} and \emph{Chandra\/} observations
that cover it. We extract the source counts from a circle with a radius of 15 arcsec and 20 arcsec in \emph{Chandra} 
and \emph{XMM-Newton} datasets respectively. The extraction regions are chosen to optimize the signal-to-noise ratio which 
correspond to an encircled energy fraction of $\goa90\%$ and $\goa75\%$ at its location in the corresponding detectors in 
\emph{Chandra} and \emph{XMM-Newton} respectively. 
The background spectrum is extracted from a nearby source-free circular region with a radius of 40 arcsec in the 
corresponding detectors. 
After the background subtraction, there are $34\pm6$ and $24\pm5$ net source counts available from the \emph{Chandra} 
and \emph{XMM-Newton} respectively. 
We compute the response files with the XMMSAS tasks RMFGEN and ARFGEN
for \emph{XMM-Newton\/} and with the CIAO tools MKACISRMF and MKARF for \emph{Chandra\/}.

With the aids of PIMMS, we can compare the count rates from different detectors. Adopting the best-fit 
spectral parameters (cf. Tab.~\ref{spec_par}), we found the count rates obtained from different detectors 
are consistent. However, as the parameters are poorly constrained, it is difficult to properly constrain 
the source variability. While the nominal observed flux is about $\sim2\times10^{-14}$~erg~cm$^{-2}$~s$^{-1}$, 
the $1\sigma$ upper limit is at the level of $\sim3\times10^{-12}$~erg~cm$^{-2}$~s$^{-1}$. Therefore, 
ascribing to the limited photon statistics of the existing data, we are not 
able to unambiguously conclude whether there is any flux variability from \xmmsrc.

\begin{figure}
  \centering
\includegraphics[width=10cm]{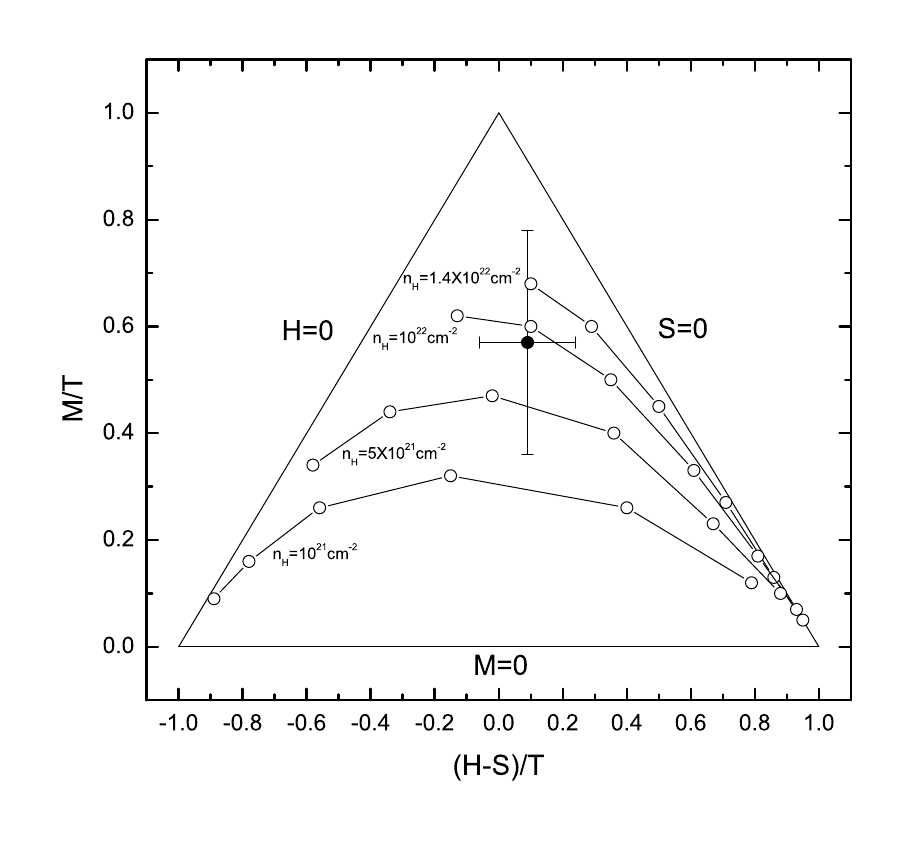}
\vspace{-1cm}
\caption{The hardness ratios of \xmmsrc\ as shown in an X-ray color-color diagram (\emph{filled circle}).
The bands are $S=0.5-1$~keV, $M=1-2$~keV and
$H=2-8$~keV. The curves within the triangular boundary are the calculated values for a power-law spectrum with photon index
varying from $\Gamma=1$ (\emph{right end}) to $\Gamma=6$ (\emph{left end}) for different adopted column densities.
Each open circle along illustrates the position of $\Gamma$ from 1 to 6 (\emph{from right to left}) in increments of 1. }
\label{ccd}
\end{figure}

To further investigate whether this source is a promising pulsar candidate, we examined its hardness ratio
and constrain its properties by means of a color-color diagram with the combined net counts obtained from 
both satellites. 
Following Elsner et al. (2008), we used three energy bands
in this analysis $S$ ($0.5-1$~keV), $M$ ($1-2$~keV) and $H$ ($2-8$~keV). 
Figure~\ref{ccd} shows the plot of $(H-S)/T$ versus $M/T$, where $T$ is the energy band $0.5-8$~keV. 
The filled circle with the $1\sigma$ error bars attached represents the
location of \xmmsrc\ in this plot. We have also computed the predicted values for a power-law spectrum 
with photon index varying from $\Gamma=1$ to $\Gamma=6$
for different values of hydrogen column absorption. The results are plotted as the curves in Figure~\ref{ccd}. 
By definition, all classes of X-ray sources should lie in the triangular boundary formed by $S=M=H=0$. 
The soft sources which lie close to the line $H=0$ are most likely field stars in the Milky Way, 
and the hard sources lie close to the line $S=0$ are likely the background active galactic nuclei (AGNs) or 
pulsars with non-thermal 
dominant X-ray emission. \xmmsrc\ is marginally located at the right side of the color-color diagram yet close to 
the center. Its hardness is found to be too hard for a field star. On the other hand, its location in the color-color 
diagram shows that it is modeled by a power-law with photon index generally larger than 3. This suggests that the 
X-ray emission of \xmmsrc\ is unlikely to be non-thermal dominant. The same inference is obtained from the spectral 
analysis (see below).

Owing to the small numbers of the collected photons, we adopt the $C-$statistic (Cash 1979) for the spectral analysis 
which is insensitive to the binning. The spectral analysis is performed with XSPEC 12.5 in the energy band of
$0.3-10$ keV and $0.5-8$ keV for the data obtained from \emph{XMM-Newton\/} and \emph{Chandra\/} respectively. 
In view of the small photon statistics, we limited our spectral analysis with simple single component model. 
The best-fit parameters of all the tested models are tabulated in Table~\ref{spec_par}. All the quoted uncertainties of 
the parameters are $1\sigma$ with 1 parameter of interest. 

We found that a single power-law 
model yields a photon index of $\Gamma_{X}=4.29^{+1.72}_{-1.09}$ which is found to be too steep to be physically 
reasonable. The large value does suggest the X-ray emission from \xmmsrc\ is rather soft which is consistent with 
the inference resulted from the aforementioned hardness analysis. 

We have also fitted the spectrum with thermal plasma models, namely MEwe-KAstra-Liedahl (MEKAL) and thermal bremsstrahlung. 
MEKAL is the code that 
models the plasma in collisional ionization equilibrium which is widely utilized to describe 
the shock-heated plasma of early-type stars 
(e.g. Sana et al. 2007; Stelzer et al. 2005). With the metal abundances fixed at solar values, the model yields 
a plasma temperature of $kT=0.51^{+0.17}_{-0.20}$~keV. However, we note that there are systematic fitting 
residuals in $\sim1-2$~keV. Repeating the analysis with metal abundances at $50\%$ and $25\%$ of the solar values, we
found the systematic residuals still present. 
Furthermore, the inferred column density is $n_{H}=1.63^{+0.63}_{-0.33}\times10^{22}$~$cm^{-2}$ 
which is found to be higher than the total Galactic neutral hydrogen absorption of $1.4\times10^{22}$~$cm^{-2}$ 
(Dickey \& Lockman 1990), unless the metal abundance is below $25\%$ of the solar values. It is possible that 
multiple temperature model might improve the fitting. Nevertheless, the limited photon statistic does not allow 
us to do so with the existing data. 
In view of its inadequacy in modeling the X-ray spectrum, we will no longer consider
this model further in this paper. 

For the thermal bremsstrahlung model, it provides a better description of the data in comparison with MEKAL. 
This model provides the description of the continuum of  
the coronal emission from a late-type star, which is presumably heated by the magnetic reconnection 
(cf. Dopita \& Sutherland 2003). It yields a plasma temperature of $kT=0.81^{+0.54}_{-0.31}$~keV. The unabsorbed flux 
inferred by the best-fit model is $1.09^{+6.98}_{-1.09}\times10^{-13}$~erg~cm$^{-2}$~s$^{-1}$ in $0.3-10$~keV. 
If the X-ray emission is indeed from a late-type star, the upper-bound of its distance is estimated to be $\sim280$~pc 
by comparing the best-fit flux to the saturated X-ray luminosity of $\sim10^{30}$~erg~s$^{-1}$ for the late-type stellar population 
(Pizzolato et al. 2003). Adopting this estimate, we convert the limiting magnitude, $m_{V}>21$, of the USNO catalog to an  
absolute magnitude of $M_{V}>14$ which suggests the star should not 
be more massive than a M-star. A deeper optical observation 
will certainly provide a key role in confirming or refuting this scenario. 

\begin{figure}
  \centering
  \includegraphics[angle=-90,width=\columnwidth]{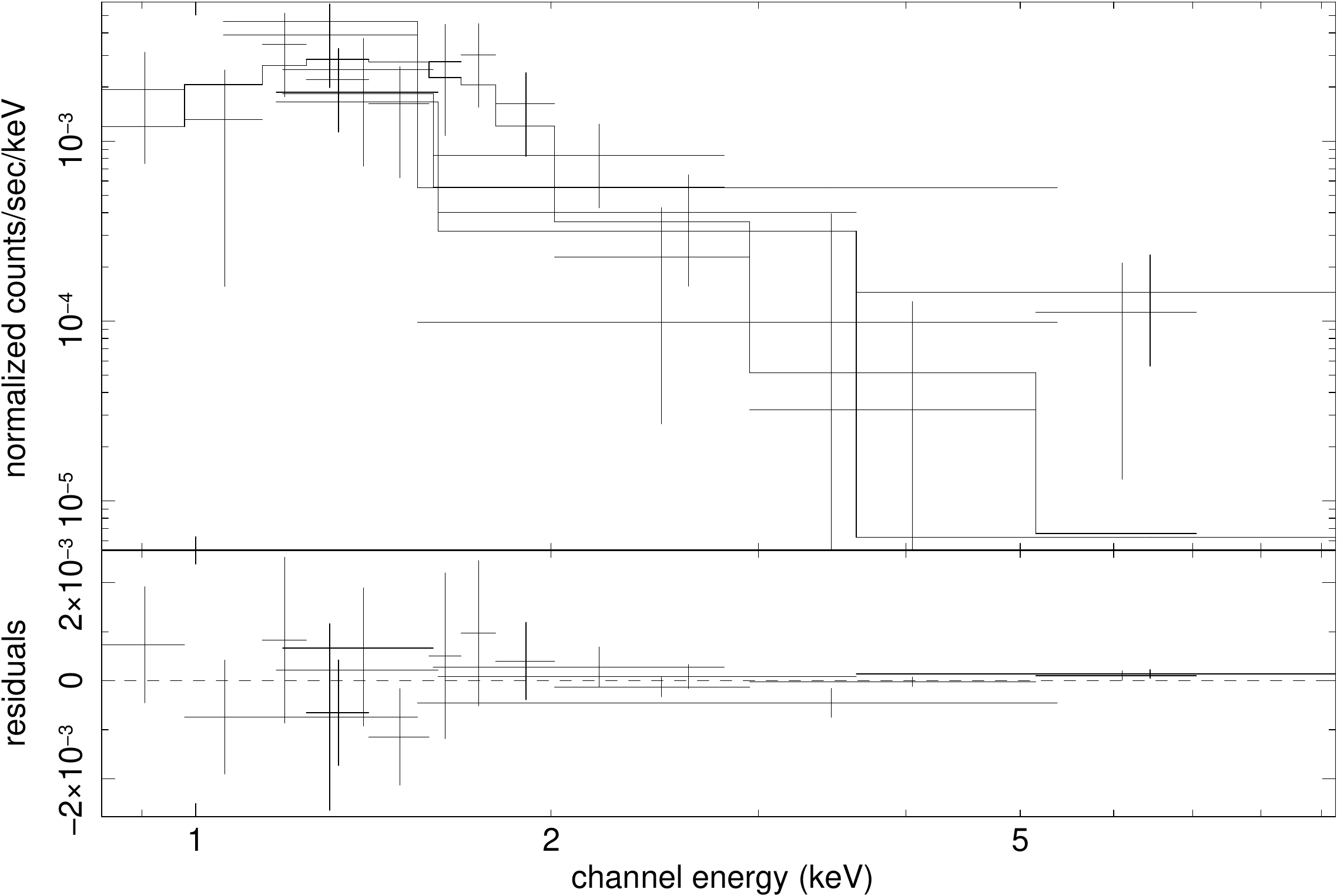}
   \caption{X-ray spectrum of \xmmsrc\ as observed with MOS1/2, PN and ACIS-I2 and
    simultaneously fitted to an absorbed blackbody model ({\it upper panel}) with
    the fitting residual ({\it lower panel}).}
\label{x_spec}
\end{figure}

For the blackbody model, assuming the source is a pulsar, it provides a description for the polar
cap heating from the return current in the outergap (Cheng \& Zhang 1999). 
The blackbody model yields an effective temperature of $kT=0.38\pm0.09$~keV. 
Interestingly, this value is well consistent with the theoretical value of the polar cap temperature 
computed by the outergap model (i.e. equation 58 in Cheng \& Zhang 1999), 
which is $kT=$0.34~keV for the rotational period and the magnetic field of \psr. 
The X-ray spectrum of \xmmsrc\ with the best-fit absorbed blackbody model is displayed in Figure~\ref{x_spec}. 
The best-fit parameters of the blackbody model yield an unabsorbed flux of 
$f_{X}=0.36^{+4.66}_{-0.32}\times10^{-13}$~erg~cm$^{-2}$~s$^{-1}$ in $0.3-10$~keV. Comparing the 
best-fit X-ray flux with the $\gamma-$ray flux (see below), 
the flux ratio is found to be $f_{X}/f_{\gamma}\sim2\times10^{-5}$ which is consistent with the typical values of $\gamma-$ray 
pulsars (e.g. Geminga). On the other hand, with the limiting magnitude of $m_{V}>21$, the nominal X-ray-to-optical flux ratio 
is $f_{X}/f_{V}>1$. This appears to be higher than that of a field star which typically has
a ratio $f_{X}/f_{\rm V}<0.3$ (Maccacaro et al. 1988). However, simply on the basis of the X-ray-to-optical flux ratio, 
the limit found for \xmmsrc\ is too low to rule out the possibility of an AGN which typically has a ratio of 
$f_{X}/f_{\rm V}<50$ (Stocke et al. 1991). Therefore, 
a deep optical observation would be important to tightly constrain its source nature. 

Although the blackbody model can yield a physically reasonable best-fit temperature and the flux for the interpretation 
of pulsar emission, one should notice that the spectral parameters and hence the fluxes are poorly constrained. Therefore, 
$f_{X}/f_{\gamma}$ and $f_{X}/f_{\rm V}$ cannot be tightly determined. Hence, we have to admit that the source nature of 
\xmmsrc\ cannot be determined unambiguously. In the spectral analysis, we are not allowed to discriminate the 
pulsar interpetation from that of a star. As the flux variability cannot be well-constrained with the existing data, we 
also cannot completely exclude the possibility of the source as an AGN. To further probe its X-ray emission nature, 
besides the aforementioned deep optical observation, a dedicated X-ray 
observation is very important in constraining the spectral properties, variability, as well as the flux ratios with respect 
to the $\gamma-$ray and optical results. 

Apart from yielding the physically reasonable temperature and the flux for the interpretation of pulsar emission, 
the blackbody model also provides relatively the least residuals among all the tested models though cannot 
be discriminated unambiguously with the small number of counts. In view of these merits, we are going to use 
the results inferred from this model for further discussion in Section 3. 

The robustness of the results quoted in this paper is checked by repeating the analysis by
incorporating the background spectrum sampled from different source-free regions. It is found that within the
$68\%$ confidence intervals the spectral parameters inferred from independent fittings are all consistent with each other.

\begin{table*}
\caption{Spectral parameters inferred from fitting the \emph{Chandra} and \emph{XMM-Newton} observed
spectra of \xmmsrc. }
\begin{center}
\begin{tabular}{lccccc}
\hline\hline
Model $^{a}$ & $C-$stat & D.O.F. & $n_{H}$ & $\Gamma_{X}^{b}$ / $kT$ & $f_{X}$ (0.3-10~keV) \\
      &                  &        & $10^{21}$ cm$^{-2}$ &  / keV  & $10^{-13}$ erg cm$^{-2}$ s$^{-1}$  \\
      \hline\\
BB   &  7.42 & 11 & $2.24^{+5.40}_{-2.24}$ & $0.38\pm0.09$  & $0.36^{+4.66}_{-0.32}$   \\           
\\
BREMSS   & 7.64  & 11  & $6.16^{+5.60}_{-3.97}$  & $0.81^{+0.54}_{-0.31}$  & $1.09^{+6.98}_{-1.09}$ \\           
\\
PL   & 8.09  & 11 & $9.99^{+8.17}_{-5.17}$  &  $4.29^{+1.72}_{-1.09}$  &  $9.21^{+184.31}_{-9.21}$ \\           
\\
MEKAL   & 8.63  & 11  & $16.33^{+6.34}_{-3.25}$  &  $0.51^{+0.17}_{-0.20}$ & $7.38^{+37.14}_{-7.38}$  \\
\\
\hline\hline
 \end{tabular}
 \end{center}
 $^{a}$   {\footnotesize PL = power-law; BB = blackbody; BREMSS = thermal bremsstrahlung; 
MEKAL = hot diffuse gas model based on Mewe et al. (1985, 1986), Kaastra (1992) and Liedahl (1995) }\\
\label{spec_par}
\end{table*}

Although the detection of X-ray pulsations at the location of \xmmsrc\ would provide us an unambiguous evidence that it is 
associated with a $\gamma-$ray pulsar, the small photon statistics and the large frame time of the existing data do not allow 
any meaningful timing analysis.

\subsection{Radio observations of \psr}

\subsubsection{Coherent radio pulsation search}

The identification of \xmmsrc\ as a possible X-ray counterpart of \psr\ 
does provide us a position with arc-second accuracy. This is 
very helpful in facilitating further multiwavelength investigations. 
With this precise position, we have searched for any coherent radio pulse emission. 

The radio pulsation search at the position of \xmmsrc\ has been 
carried out by using the 25-m radio telescope at Nanshan, operated by
Urumqi Astronomical Observatory (UAO). 
The observing system has a dual-channel cryogenic receiver that receives
orthogonal linear polarizations at 18~cm.  
After mixing down to an intermediate
frequency, the two polarizations are each fed into a filter
bank of 128 contiguous channels, each of width 2.5
MHz. The outputs from the channels are then square-law
detected, filtered and one-bit sampled at 0.5 ms interval.
The data streams of all 256 channels are written to disk for
subsequent off-line processing.
For more details about this system, please refer to Wang et al. (2001). 

In our observation, we did not find any convincing signal and we placed 
an upper-limit for any pulsed radio emission of 0.1~mJy at the position of 
\xmmsrc. 

\subsubsection{Search for radio emission feature from NRAO/VLA Sky Survey (NVSS)}

\begin{figure}
\centering
 \includegraphics[width=9.3cm]{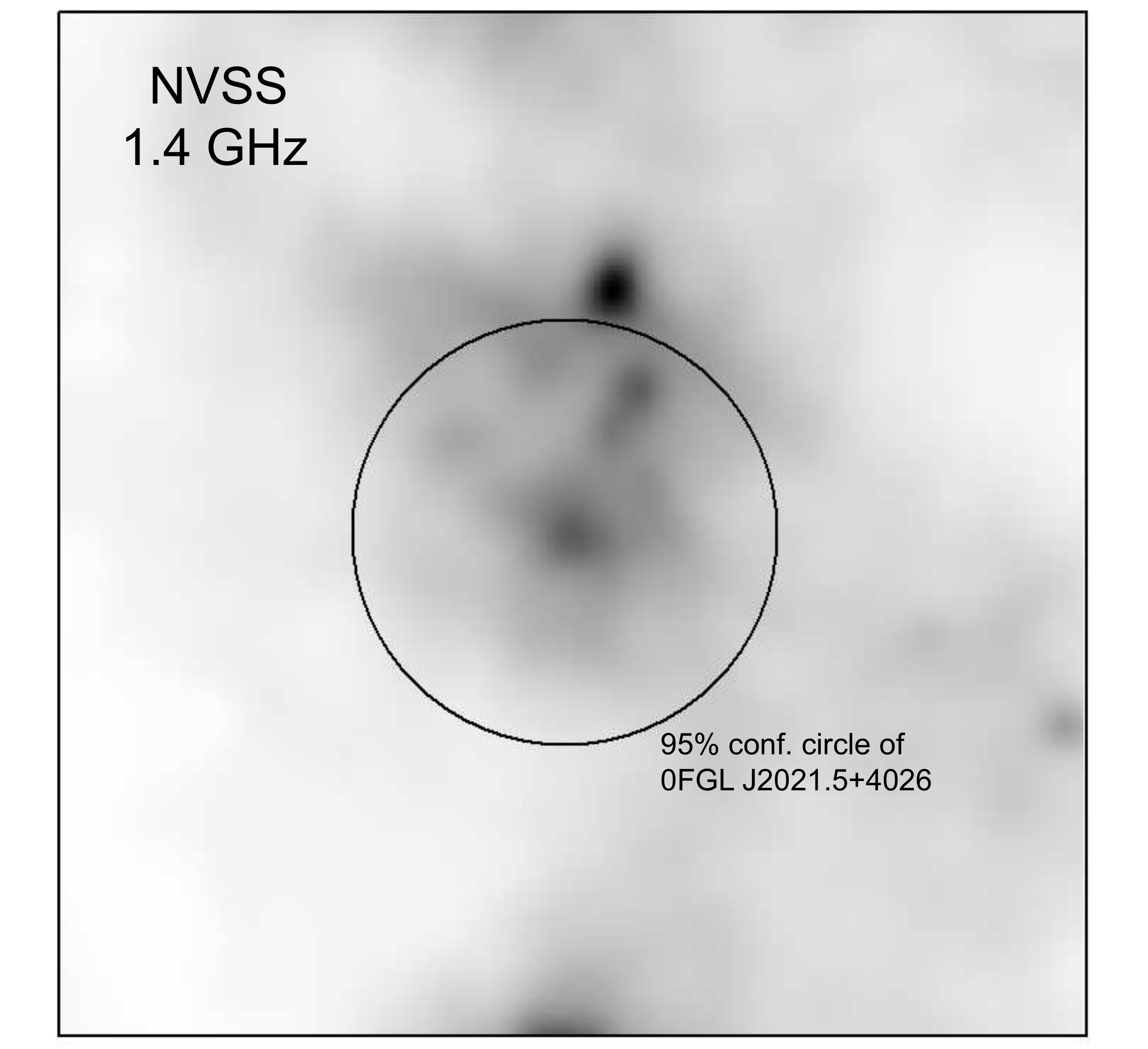}
  \caption[]{The 1.4 GHz NVSS image of a $15\times15$ arcmin$^{2}$ field centered at the nominal $\gamma-$ray position
  of 0FGL~J2021.5+4026. The $95\%$ error circle of the $\gamma-$ray source is illustrated.}
  \label{nvss1}
\end{figure}

\begin{figure}
\centering
 \includegraphics[width=9.3cm]{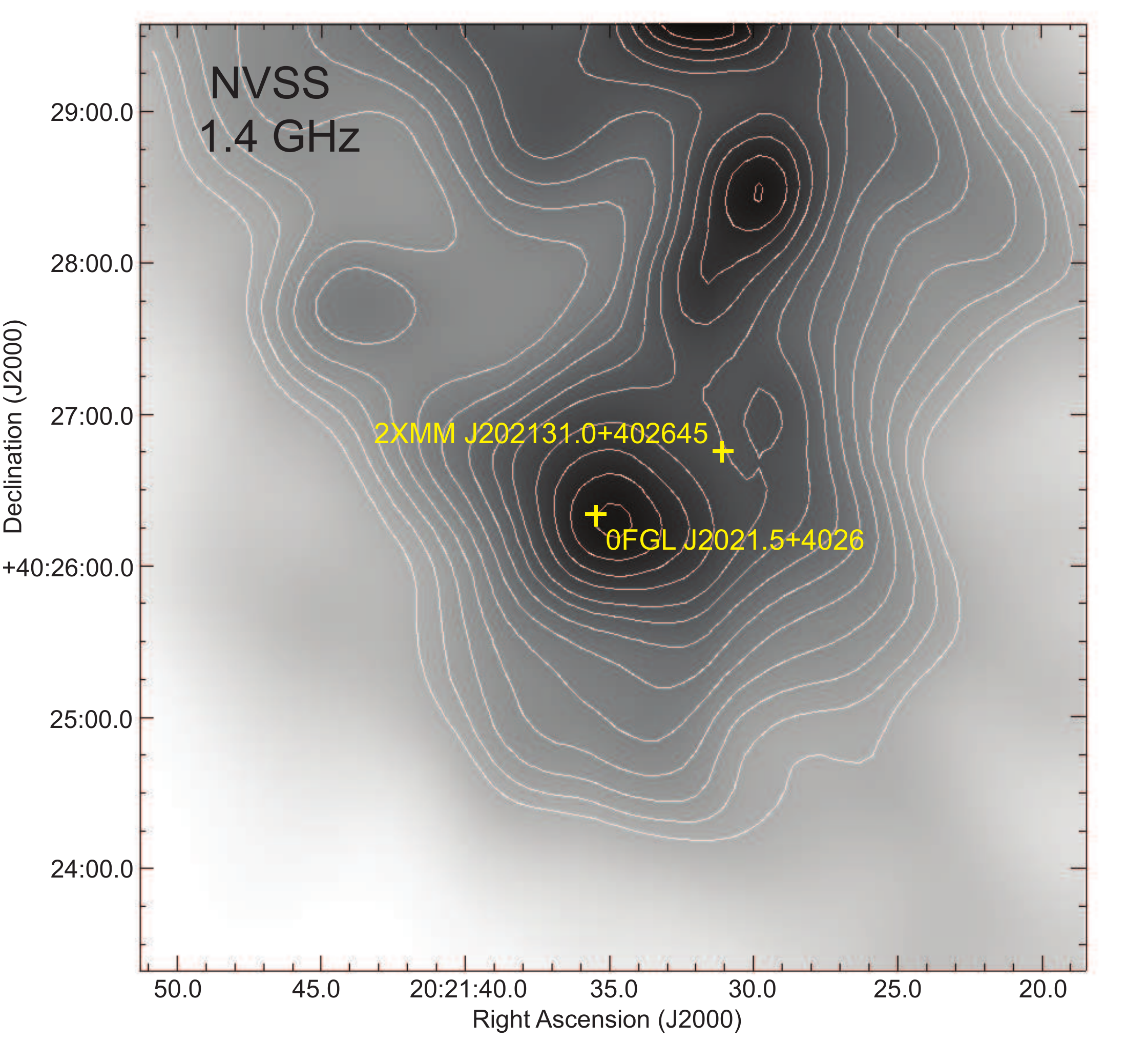}
  \caption[]{A $6\times6$ arcmin$^{2}$ close-up view of Figure~\ref{nvss1}. The field-of-view of this image corresponds 
 to the size of the $\gamma-$ray error circle. 
  The position of \xmmsrc\ is also indicated. The radio contours are at the levels between $10-25$ mJy/beam.}
  \label{nvss2}
\end{figure}

We have also searched for any radio counterpart for \psr\ with the data from the NVSS database. 
(Condon et al. 1998). Interestingly, we have identified radio excesses within the $\gamma-$ray error circle of \psr\ 
(see Figure~\ref{nvss1}). 
A $6\times6$~arcmin$^{2}$ close-up view centered on the nominal 
$\gamma-$ray position of \psr\ reported in Abdo et al. (2009a) is displayed in Figure~\ref{nvss2}. 
Radio contours calculated at 
the levels between $10-25$~mJy/beam are overlaid. 
We have identified a feature with a size of about
3 arcmin$\times$1.5 arcmin in the center of this radio map. 
The peak of this radio feature is found to be at the south-east from the position of \xmmsrc. 
Apart from the aforementioned feature, another radio excess extends for $\sim3$ arcmin from \xmmsrc\ to the north-west. 
Adopting the FWHM of the beam and the rms fluctuation of the image of 45 arcsec and 0.45 mJy/beam respectively
(cf. Condon et al. 1998), we estimated the flux densities at 1.4~GHz of the southeastern and the northwestern features to
be $139\pm4$~mJy and $85\pm2$~mJy respectively. These correspond to $(5.84\pm0.17)\times10^{-17}$~ergs~cm$^{-2}$~s$^{-1}$ and 
$(3.57\pm0.08)\times10^{-17}$~ergs~cm$^{-2}$~s$^{-1}$ for an effective bandwidth of 42~MHz respectively. 

It is interesting to notice that \xmmsrc\ is located approximately in between this two features. If such alignment is 
confirmed, this will suggest a possible bipolar outflow from the pulsar. 
Unfortunately, the limited angular resolution of NVSS data does not allow us to conclude this possible alignment. 
Future observation with the dedicated high resolution aperture synthesis by VLA can help us to confirm (or refute) this 
suggested scenario. 

\subsection{$\gamma-$ray analysis of \psr} 

With the already publicly available data of the \emph{FERMI} 
$\gamma-$ray LAT All-Sky survey we have carried out an analysis
of \psr\ centered at the accurate X-ray position derived by analyzing 
\emph{Chandra} data (see Sec 2.1). We have studied its 
$\gamma-$ray spectral and temporal properties in details. In order to 
get the most significant results for the spectral analysis we
took all available events from the start of the LAT All-Sky survey 
4 August 2008 until 26 September 2009, 
which leads to a total effective exposure time of 13246 ksec as the 
observatory scans the entire sky once every three hours. Events have been
chosen in the energy range  of 100 MeV to 300 GeV as the effective 
area is changing rapidly with energy at less then
100 MeV and because of residual uncertainties in the instrument 
response (Porter 2009). For general filtering purposes of the
extracted data we used the LAT Pass6 Version 3(P6\_V3) instrument 
response functions as we are interested in
point source analysis. We extracted the events 
within a circle of $2.5^{\circ}$ radius centered at \xmmsrc\ in 
order to minimize the contamination by other sources and kept the photon containment 
fraction $\ga95\%$ for photons with energies $\ga0.5$~GeV. We have further 
excluded the events with a zenith angle larger than 105$^{\circ}$ as recommended
by the standard filtering of the \emph{FERMI} data group
\footnote{Please refer to http://fermi.gsfc.nasa.gov/ssc/ for 
further details.}. This is due to the fact that the maximum zenith angle selection is
applied to exclude time periods when the region of interest is too close to 
the Earth's limb, resulting in elevated background levels. We finally ensured
that only diffuse class photons were used by setting \emph{event\_class=3}. 
Good time intervals have been chosen by using the 
spacecraft data file to ignore all events acquired during the passage of 
the South Atlantic Anomaly and allow only \emph{data\_qual=1} events. 
After filtering for diffuse background emission there are 22162 total source counts available. 

As \psr\ is the brightest of the 16 newly found pulsars by LAT we were 
able to study its spectrum with superior photon statistic (Abdo et al. 2009a). 
We notice that the data above 10 GeV are highly fluctuated which  
can possibly be due to the fact of a non-negligible background
contamination from charged particles.\footnote{cf. http://fermi.gsfc.nasa.gov/ssc/data/analysis/LAT$\_$caveats.html}  
With these considerations, we choose an energy range of $0.1-10$~GeV for the spectral analysis. 
The response file was computed using the \emph{gtrspgen} task and 
the spectrum was grouped with a binning of 100 logarithmically uniform energy bins. 

\begin{figure}
 \centering
  \includegraphics[angle=-90,width=\columnwidth]{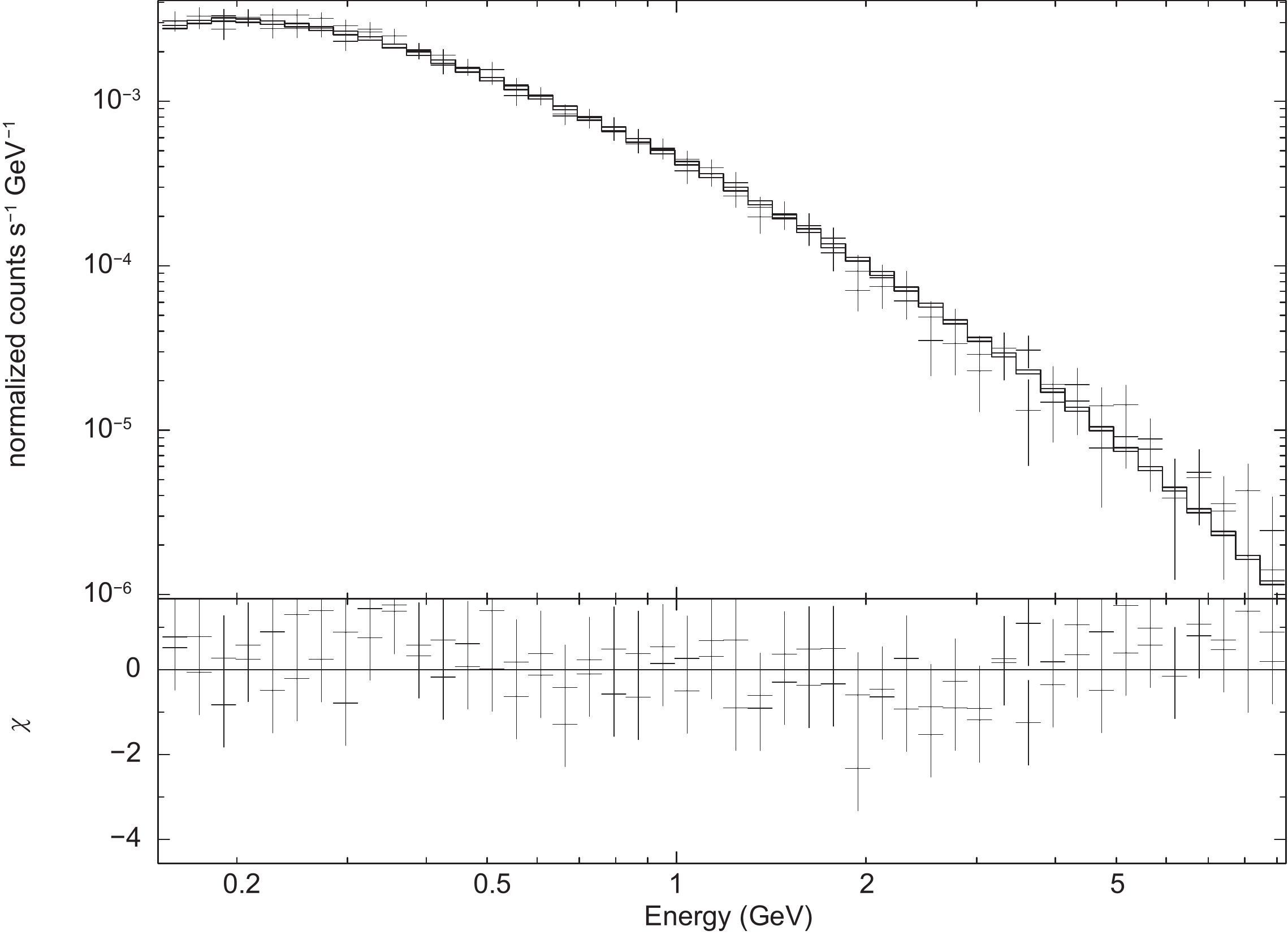}
  \caption{$\gamma-$ray spectrum of \psr\ as observed with LAT and fitted with an exponentially 
     cut-off power-law model ({\it upper panel}) and the fitting residuals ({\it lower panel}).} 
\label{g_spec}
\end{figure}

The $\gamma-$ray spectrum of \psr\ can be well-described by an exponentially cut-off model with $\chi^{2}_{\nu}$ of 0.9997 
for 97 degrees of freedom (see Figure~\ref{g_spec}). The best-fit model yields a photon index of 
$\Gamma_{\gamma}= 1.85^{+0.03}_{-0.04}$ with a cut-off energy $E_{C}=3.86^{+0.58}_{-0.48}$ GeV.
The given errors are at the 1$\sigma$ confidence level for single parameter. 

\begin{figure}
 \centering
 \includegraphics[angle=-90,width=\columnwidth]{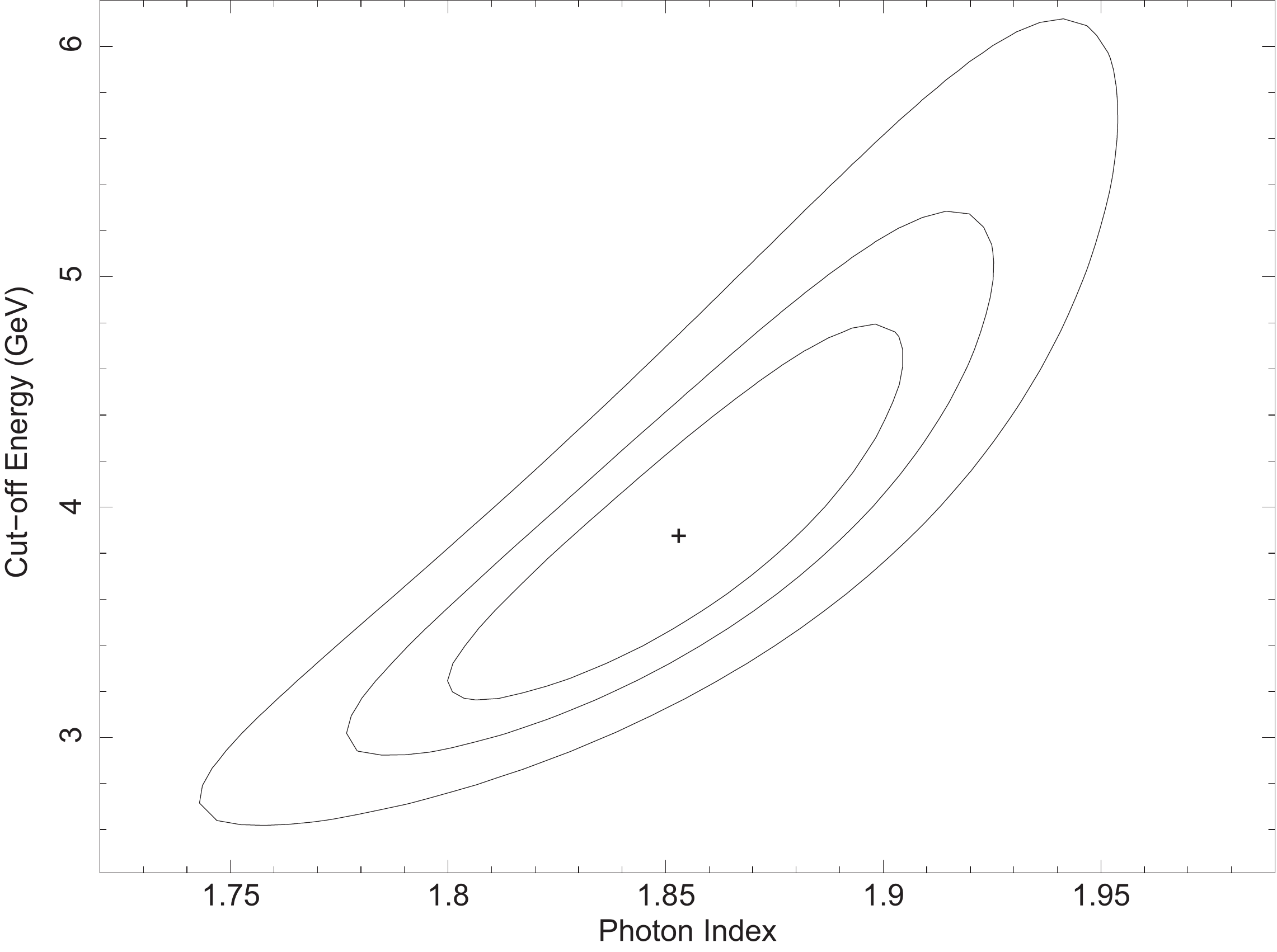}
  \caption{1$\sigma$, 2$\sigma$ and 3$\sigma$ confidence contours for the cuf-off 
    powerlaw model fitted to the $\gamma-$ray spectrum of \psr\ }
\label{gamma_contour}
\end{figure}

We have also computed the error contours to demonstrate the relative parameter
dependency of the photon index versus the cut-off energy and plotted this
in Figure~\ref{gamma_contour}. 
Taking these parameters into consideration the 
$\gamma-$ray flux is found to be $f_{\gamma}=(1.45^{+2.30}_{-0.88})\times10^{-9}$ erg cm$^{-2}$ s$^{-1}$ 
in the range of $0.1-10$~GeV. The spectral properties inferred from the whole-year LAT data are 
consistent with those reported by Abdo et al. (2009c) within $1\sigma$.  

\begin{figure}
 \centering
 \includegraphics[width=\columnwidth]{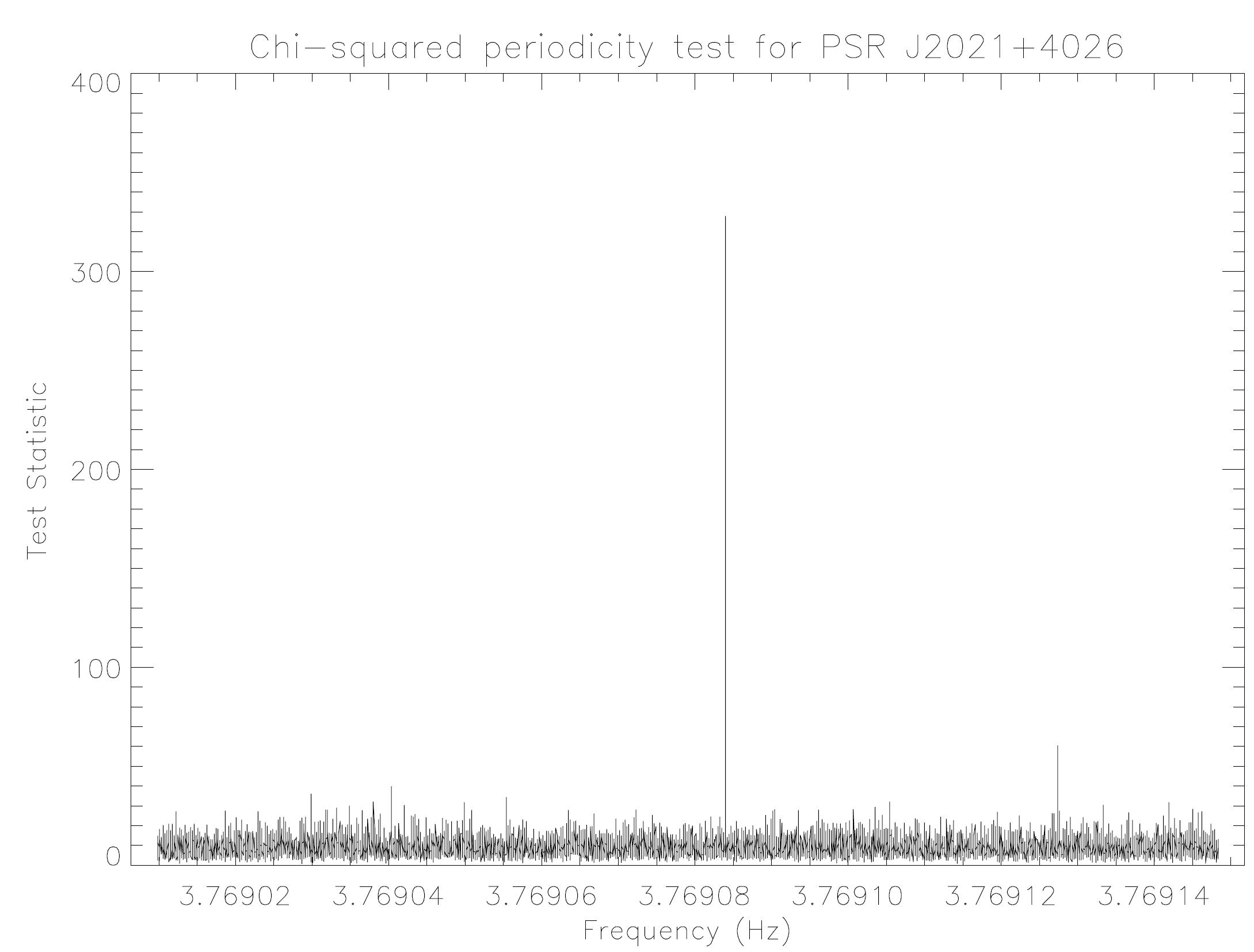}
  \caption{$\chi^{2}$test statistic periodogram for \psr\ showing the most probable frequency of 3.76908 Hz}
\label{pow_spec}
\end{figure}

To further investigate the nature of \psr, we have independently 
performed a temporal analysis. The results of this analysis will be 
further taken as inputs for a modeling in the context of outergap model in \S3. 
We have iterated the temporal analysis with different sizes of extraction region. 
We found that the signal-to-noise ratio of the pulse profile is optimized 
with an extraction radius of 1$^{\circ}$, therefore we adopted this region 
for the subsequent analysis. 
The retrieved event file was then processed in the same way 
as aforementioned with events in the energy range of 0.1 GeV - 300 GeV. 
For the barycentric correction, we adopt the precise position provided by \xmmsrc. For the
pulsation search the \emph{gtpsearch} task of the \emph{FERMI} analysis 
software was used with the $\chi^{2}$ test statistics.
We used 10 phase bins for the $\chi^{2}$ test
with a step size of 0.5 in units of Fourier resolution that lead us 
to a number of 5000 independent trials. By checking a frequency range from
3.76885824639721 to 3.76929999373175 Hz and using $\dot{f}=-7.78\times10^{-13}$~Hz~s$^{-1}$ 
we found the most probable frequency to be 3.76908389(3)~Hz 
at a test statistic of $\chi^{2}=328.0$ with 9 degree of freedom (see Figure~\ref{pow_spec}).  
The number in the parenthesis is the uncertainty of the last digit of the quoted frequency which 
corresponds to the Fourier resolution. 
We have also performed the analysis by combining $Z^{2}_{n}$ test and 
$H-$test, where $n$ is the numbers of harmonics (Buccheri et al.~1983; 
De Jager, Swanepoel, \& Raubenheimer 1989). With this independent analysis, we reported the 
periodic signal with a frequency equal to the aforementioned. Using the $H$-test, we found that 
$H$ is maximized for 2 harmonics. The calculated $Z^{2}_{2}$ is 309.7 which implies a nominal 
chance probability of $4\times10^{-62}$.  

We have also repeated the analysis with different positions within the $95\%$ $\gamma-$ray error circle 
adopted for barycentric correction. Among all the tested positions, the X-ray position of \xmmsrc\ results in 
the best test statistics. For example, the nominal $\gamma-$ray position provided by Abdo et al. (2009a) results 
in a $\chi^{2}$ of 280.9 which is lower than that resulted from adopting the X-ray position. 
This provides another support for the possible association between \xmmsrc\ and 
\psr. 

\begin{figure}
 \centering
 \includegraphics[width=9.5cm]{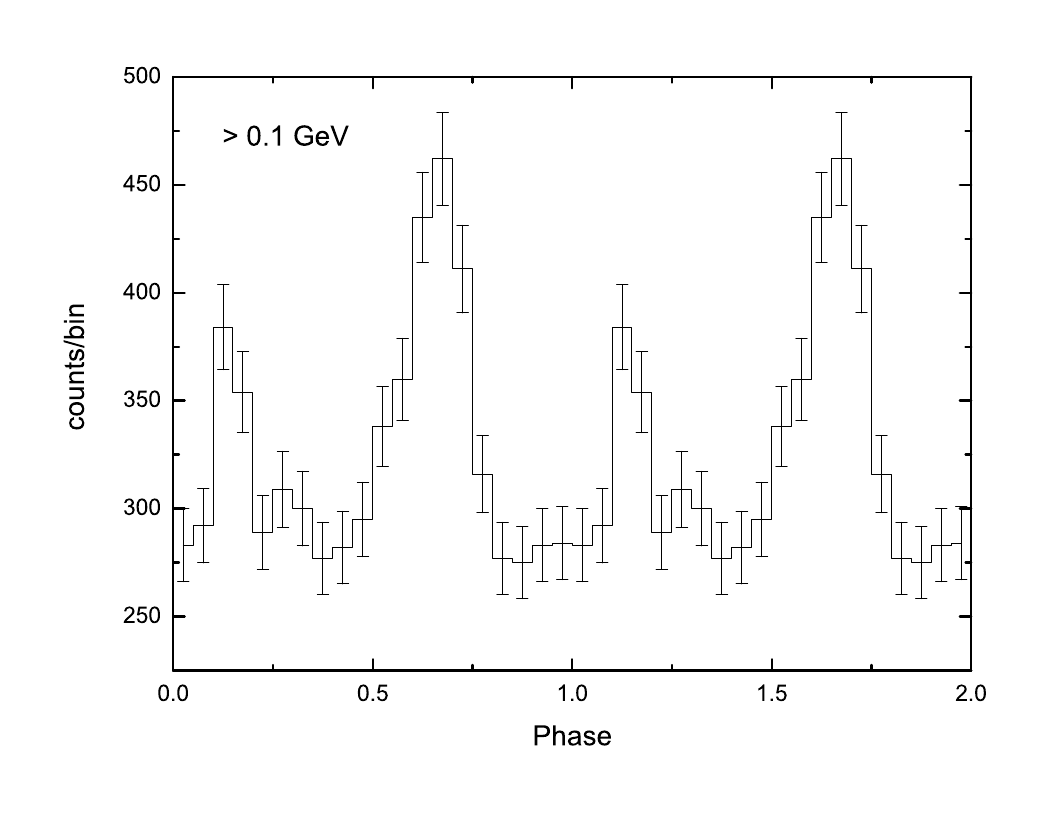}
  \caption{The $\gamma-$ray phase folded light curve of \psr\ as observed by LAT in the range of 0.1 GeV -300 GeV. }
\label{lat_pulse}
\end{figure}

The $\gamma-$ray pulse profile folded at the aforementioned period is shown in Figure~\ref{lat_pulse}. 
We found that \psr\ has a double peaked light curve with the peak separation of 
$162^{\circ}$ or $198^{\circ}$, depending on which peak is leading. In computing the pulsed fraction of the resultant 
light curve, we found that about $\sim$ 54\% of the collected photons are pulsed. 
This light curve provides us a crucial input for 
modeling the emission geometry of this pulsar (see below).

With the data collected from a period somewhat longer than one year, it is also instructive to investigated if there is 
any unusual spin-down behavior of \psr, such as glitches. We have divided the data into segments with equal time-span 
and analyzed these sub-datasets independently. Our inferred spin-down rate is found to be steady and 
fully consistent with the value reported by Abdo et al. (2009c). Therefore, we concluded that there is no evidence for any 
unusual spin-down behavior of \psr\ in the last year. Therefore, the frequency derivative is fixed at 
$-7.78\times10^{-13}$~Hz~s$^{-1}$ throughout the aforementioned analysis. 

\section{Discussion}
\begin{figure}
 \centering
 \includegraphics[width=9.5cm]{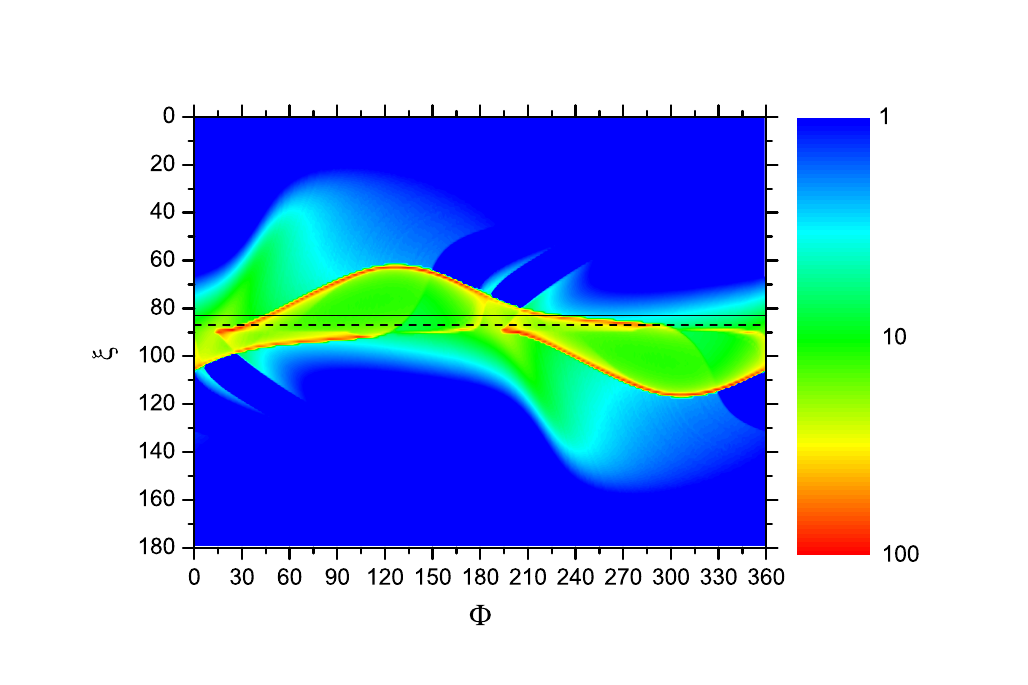}
  \caption{The phase-plot diagram of the  emissions for 
the inclination angle of $\alpha=40^{\circ}$. The color represents the 
intensity of the emissions (arbitrary unit). The solid and dashed
 horizontal lines are corresponding to the viewing angle of $\xi=83^{\circ}$ and 
$\xi=87^{\circ}$, respectively.}
\label{J2021_40}
\end{figure}

\begin{figure}
 \centering
 \includegraphics[width=9.5cm]{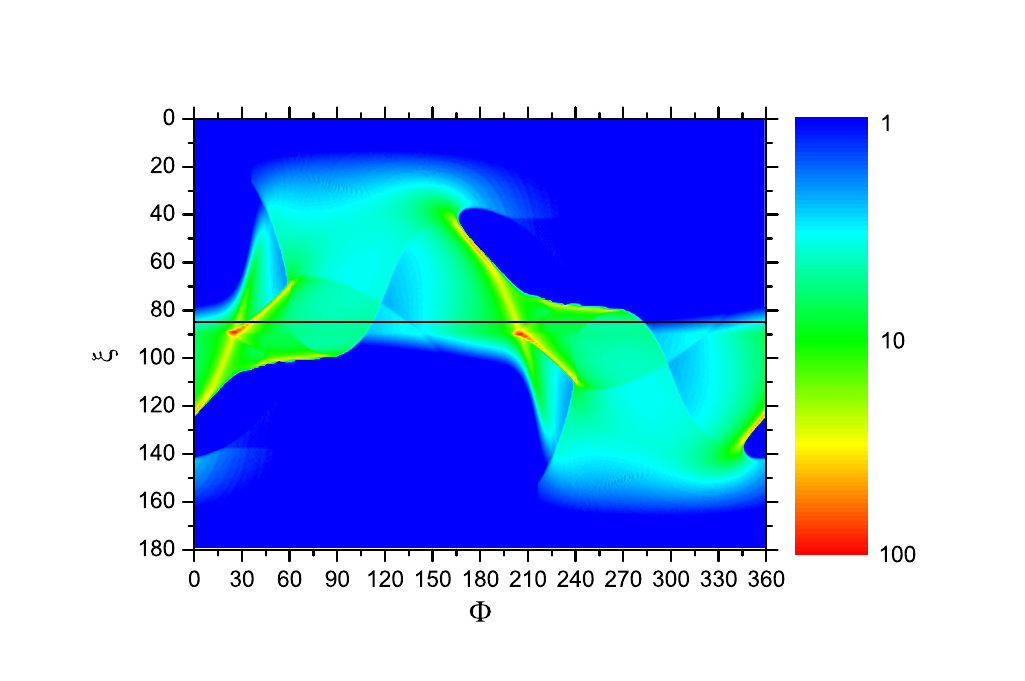} 
  \caption{The phase-plot diagram of the  emissions for 
the inclination angle of $\alpha=60^{\circ}$. The color represents the 
intensity of the emissions (arbitrary unit). The solid 
 horizontal line is corresponding to the viewing angle of $\xi=85^{\circ}$.}
\label{J2021_60}
\end{figure}
\subsection{Viewing geometry}
Figures~\ref{J2021_40} and~\ref{J2021_60}
represent the phase-plots of the $\gamma$-ray emissions from the outer
gap accelerator (e.g. Tang et al. 2008) for the inclination of 
$\alpha=40^{\circ}$ and $60^{\circ}$, respectively. The color represents the 
intensity of the emissions. In the model, we assume that the outer gap
extends between the radial distance of ten times the stellar radius
and the light radius.

The $\gamma$-ray emission from \psr\ is observed with  double peak
structure with a wide phase separation between the two peaks 
(Figure~\ref{lat_pulse}). On the other hand,
because of the lack of radio pulses from \psr,   there is an ambiguity on the
phase separation of the two peaks in $\gamma$-ray emission, that is, 
$162^{\circ}$ or $198^{\circ}$, depending on which peak is leading. 
With the phase-separation of $162^{\circ}$ or $198^{\circ}$, 
the outer gap accelerator model predicts the viewing angle close to 
$\xi\sim90^{\circ}$. 
For example, if we assume the inclination
angle of $\alpha=40^{\circ}$, the viewing angles of $\xi$ 
(solid horizontal line in Figure~\ref{J2021_40}) and $\xi$ (dashed horizontal line) 
provide the phase-separation 
 of $168^{\circ}$ and $198^{\circ}$, respectively. For the inclination
 angle $\alpha=60^{\circ}$,  the phase-separation of 
$168^{\circ}$ can be explained by the viewing angle of $\xi$, while the
phase-separation of $198^{\circ}$ can not be produced by any viewing
angle. Although the observed large phase-separation can constrain  
the viewing angle close to $\xi\sim 90^{\circ}$, a wide range of the
viewing angle is allowed.  
To resolve the  ambiguity on the inclination angle, it will require a
more detailed study (in particular the spectrum) 
 of the $\gamma$-ray emission, which will be a purpose of 
 the subsequent theoretical paper.

\subsection{Association among  \psr,  \xmmsrc\ and SNR G78.2+2.1}
\label{association}
\psr\ is a very bright newly uncovered $\gamma-$ray pulsar. The
non-detection of any radio 
pulsed signal and its high energy emission properties suggest the pulsar 
to be another example of Geminga-like pulsars. 
Recent observations with the \emph{Milagros} $\gamma-$ray Observatory
have found a $4.2\sigma$ TeV excess in the error circle of \psr\
(Abdo et al. 2009e). The reported TeV flux of this feature is 
$(35.8\pm8.5)\times10^{-17}$~TeV$^{-1}$~cm$^{-2}$~s$^{-1}$. 
The presence of the multiwavelength features from radio to TeV regime
have further made it an interesting target. 
\psr\ and X-ray source, 2XMM J202131.+402645 
 can be observed at  vicinity of the edge of the  
 supernova remnant (SNR) G78.2+2.1, which
 may be the origin of the TeV emission. 
  In the following, we will argue that the present  results of  the 
 X-ray and $\gamma$-ray
analysis indicate an  association among \psr, 2XMM J202131.+402645 
and  SNR  G78.2+2.1. 

First we discuss the ages of SNR G78.2+2.1 and \psr. 
Using ASCA observations,   Uchiyama et al. (2002) deduced the shock
velocity of the shell of SNR G78.2+2.1 to be 800~$\mathrm{km s^{-1 }}$.
 Given the angular size of $\theta\sim 30'$ 
and distance $d\sim 1.5\pm 0.5$~kpc (Landecker, Roger \& Higgs 1980), 
the adiabatic age gives 6.6(d/1.5kpc)~kyr, 
which is in  agreement with the age deduced from
 the  optical observations (Mavromatakis 2003). We find that 
the age of SNR G78.2+2.1  is about one order  magnitude  smaller
 than the spin-down age of $\tau\sim 77$~kyr for PSR J2021+4026.  
However, the discrepancy between the real age and the spin down age may
be expected, if   PSR J2021+4206  was born with a spin period
 close to  current one. It can be expected that 
PSR J2021+4026 is much younger than the age inferred from
 the spin down age, such as  PSR J0538+2817, which 
 has a spin down age of 
620~kyr, but its true age is 40~kyr (Ng et al. 2007). Therefore, the
discrepancy between  the age of SNR G78.2+2.1 and the spin down age
 of PSR J2021+4026 does not imply that   
\psr\ and the supernova remnant G78.2+2.1 are not associated with each other.

In fact, we expect that SNR G78.2+2.1 and \psr\ are associated with each
other, as follows. First, adopting  the distance $d\sim 1$~kpc to \psr, the
efficiency, $\eta$, which is defined by the ratio between $\gamma$-ray 
luminosity and spin down luminosity, is provided as 
$\eta\sim 0.1\delta\Omega (d/\mathrm{kpc})^2$, where $\delta\Omega$ is 
the solid angle.  This large efficiency with the distance 
 is consistent  with the typical value of
the efficiency of the middle spin-down 
age pulsars such as Geminga, which has $\eta\sim
0.07\delta\Omega$. 
Therefore,  the distance to the SNR G78.2+2.1 provides a consistent
 efficiency with the spin down age.  Secondly, \psr\ is located 
$\sim 7.8$ arcmin off-axis from the geometrical center of SNR G78.2+2.1. 
Assuming that off-axis of the location of \psr\ is caused by 
the proper motion in the space, the space  velocity of PSR J2021+4026 is
estimated to be  $v_p\sim 340 (d/1\mathrm{kpc})$km/s,
 which is a typical velocity of observed pulsars (Hansen \& Phinney 1997; Hobbs et al. 2005). 
On these ground, the association between \psr\ and SNR G78.2+2.1 is more likely. 

Finally, we briefly discuss the association between  2XMM J202131.+402645 and 
\psr. First, only 2XMM J202131.+402645 is a persistent and relatively
bright $X$-ray source located within the 
 95\% $\gamma-$ray error circle of \psr\ (see section~2). Second,  
the nominal X-ray flux in 2-10~keV, $f_{X}\sim 
8\times 10^{-15}~\mathrm{erg/cm^2/s}$, of  2XMM
J202131.+402645  provides a X-ray luminosity  of $L_{X}\sim
 10^{30}f_{\Omega}~\mathrm{erg/s}$, where $f_{\Omega}$ is the solid
 angle divided by $4\pi$.  Comparing the X-ray luminosity of other
 pulsars (e.g. Possenti et al. 2002), we find that the X-ray luminosity 
 $L_{X}\sim 10^{30}f_{\Omega}~\mathrm{erg/s}$ of 2XMM J202131.+402645
 is consistent with typical values of the pulsars which have a spin
 down luminosity similar to that of \psr,  $\dot{E}\sim 10^{35}$~erg/s.
 It is also found that the ratio of the X-ray
 flux of 2XMM J202131.+402645 deduced from the best-fit blackbody model and $\gamma$-ray flux of
 \psr, $f_X/f_{\gamma}\sim2\times10^{-5}$, 
 is consistent with typical values of $\gamma$-ray pulsars with a middle spin-down age like Geminga.  
 On these ground, we suggest that 2XMM~J202131.+402645 
is the plausible X-ray counter part of \psr. 

Although the interpetation of the pulsar nature is tempting, we have to emphasize that small photon 
statistics of the existing data does not allow us to confirm this unambiguously. Specifically, we
cannot tightly constrain the aforementioned flux ratios, spectral parameters as well as the variability. 
Therefore, we have to admit that we cannot exclude the possible nature of \xmmsrc\ as a star or an AGN. 
Dedicated X-ray and optical observations are most important in discriminating these competing X-ray emission 
scenarios. Obviously, if X-ray pulsations that consistent with the rotational period of \psr\ can be 
detected in the future, this will certainly provide the most decisive nature of \xmmsrc. 

\section{Summary}
In this study, we have investigated the multiwavelength emission nature of \psr\ in 
details. Searching for the X-ray counterparts of this new and bright $\gamma-$ray pulsar, 
we have identified \xmmsrc\ as the promising candidate. 
We have also examined the $\gamma-$ray data collected by FERMI LAT with an exposure somewhat more than one 
year and tightly constrained its spectral and temporal properties in MeV$-$GeV regime. 
We found that the X-ray position of \xmmsrc\ is consistent with that of the optimal $\gamma-ray$ timing solution. 
We have further modeled 
the $\gamma-$ray light curve in the context of outer gap accelerator model and provided constraints on its 
emission geometry. The nominal X-ray--to--$\gamma$-ray flux ratio of \psr\ is found to resemble that of Geminga. 
Furthermore, if \psr\ was born with a spin period close to the current one, it is likely to be born in the explosion that 
created SNR G78.2+2.1 and has a projected kick velocity of few hundred km/s which is typical for the known pulsar 
population. At the distance of SNR G78.2+2.1, the conversion efficiencies in $\gamma-$ray and X-ray of this pulsar 
are both found to be consistent with those of Geminga. Together with the non-detection of any pulsed radio signals, 
the high energy emission properties of \psr\ suggest it to be a new member of Geminga-like pulsars. 

\section*{Acknowledgments}

LT would like to thank DFG for financial support in SFB TR 7 Gravitational Wave Astronomy and the members of the 
neutron star group at AIU for fruitful discussions and useful comments.
KSC is supported by a GRF grant of Hong Kong Government under HKU700908P. The authors would like to thank 
C.Y. Ng for useful discussion.

\label{lastpage}

\vspace{-0.5cm}
\begin{thebibliography}{}

  \bibitem[]{} Abdo, A. A. et al., 2009a, Science, 325, 840

  \bibitem[]{} Abdo, A. A. et al., 2009b, Science, 325, 848

 \bibitem[]{} Abdo, A. A. et al., 2009c, submitted to ApJS (arXiv:0910.1608v2)

  \bibitem[]{} Abdo, A. A. et al., 2009d, ApJS, 183, 46  

  \bibitem[]{} Abdo, A. A. et al. 2009e, ApJ, 700, L127 

  \bibitem[]{} Becker, W. et al., 2004, ApJ, 615, 897

  \bibitem{} Buccheri, R. et al., 1983, A\&A, 128, 245

  \bibitem[]{} Cash, W., 1979, ApJ, 229, 939

  \bibitem[]{} Cheng, K. S., 2008, in W.~Becker ed., Astrophysics
   and Space Science Library, Neutron Stars and Pulsars, Springer, Berlin, p.481

   \bibitem[]{} Cheng, K. S., \& Zhang, L. 1999, ApJ, 515, 337

   \bibitem[1998]{condon} Condon, J. J. et al., 1998,
      AJ, 115, 1693

   \bibitem{} De Jager, O. C., Swanepoel, J. W. H., \& Raubenheimer, B. C. 1989, A\&A, 221, 180

   \bibitem[1990]{dickey} Dickey, J. M., \& Lockmann, F. J., 1990,
      ARA\&A, 28, 215

   \bibitem{} Dopita, M. A., \& Sutherland, R. S. 2003, Astronomy and Astrophysics Library
	Astrophysics of the Diffuse Universe, Springer, Berlin

  \bibitem[]{} Elsner, R. F. et al., 2008, ApJ, 687, 1019

   \bibitem[]{} Green D.A., 2009, A Catalogue of Galactic Supernova Remnants (2009 March version).
    Mullard Radio Astronomy Observatory, Cambridge

  \bibitem[1997]{hansen}Hansen, B. M. S.; Phinney, E. S.,  1997, MNRAS, 
291, 569


   \bibitem[]{} Harding, A. K., 2008, in W.~Becker ed., Astrophysics
      and Space Science Library, Neutron Stars and Pulsars, Springer, Berlin, p.521

   \bibitem[1999]{hartman} Hartman, R. C. et al., 1999,
      ApJS, 123, 79

    \bibitem[]{} Hobbs, G., Lorimer, D. R., Lyne, A. G., \& Kramer, M. 2005, MNRAS, 360, 974 




   \bibitem[]{} Hui, C. Y., \& Becker, W. 2009, A\&A, 494, 1005

\bibitem[]{} Kaastra, J.S. 1992, An X-Ray Spectral Code for Optically Thin Plasmas,
Internal SRON-Leiden Report, updated version 2.0

 \bibitem[1980]{Landecker} 	
	Landecker, T. L.; Roger, R. S.; Higgs, L. A., 1980, A\&AS, 39,133 

\bibitem[]{} Liedahl, D.A., Osterheld, A.L., and Goldstein, W.H. 1995, ApJ, 438, L115 

\bibitem[]{}
Maccacaro, T., Gioia, I. M., Wolter, A., Zamorani, G., \& Stocke, J. T. 1988, ApJ, 326, 680 

\bibitem[2003]{mavromatakis}Mavromatakis, F., 2003, A\&A, 408, 237

\bibitem[]{} Mewe, R., Gronenschild, E.H.B.M., and van den Oord, G.H.J. 1985,A\&AS, 62, 197

\bibitem[]{} Mewe, R., Lemen, J.R., and van den Oord, G.H.J. 1986, A\&AS, 65, 511 
	    
   \bibitem[]{} Monet, D., et al. 2003, AJ, 125, 984

\bibitem[2007]{Ng}Ng, C.-Y.; Romani, R. W.; Brisken, W. F.; Chatterjee,
	    S; Kramer, M.,  2007, ApJ, 654, 487

\bibitem[]{} 	
Pizzolato, N., Maggio, A., Micela, G., Sciortino, S., \& Ventura, P. 2003, A\&A, 397, 147 

\bibitem[2002]{possenti} Possenti, A.; Cerutti, R.; Colpi, M.;
	    Mereghetti, S.,  2002, A\&A, 387, 993

   \bibitem[]{} Porter, T. A., 2009, arXiv:0907.0294v1

\bibitem[]{} Sana, H. et al., 2007, in St. Louis N., Moffat A. F. J., eds, ASP Conf. Ser. Vol. 367, 
             Massive Stars in Interactive Binaries. Astron. Soc. Pac., San Francisco, p.109 
  
\bibitem[]{} Stelzer, B. et al., 2005, ApJS, 160, 557  

\bibitem[]{}
Stocke, J. T., Morris, S. L., Gioia, I. M., Maccacaro, T., Schild, R., Wolter, A., Fleming, T. A., \& Henry, J. P. 1991,
ApJS, 76, 813 

\bibitem[2008]{Tang}	
	Tang, Anisia P. S.; Takata, J.; Jia, J.J.; Cheng, K. S, 
	2008, ApJ, 676, 562

\bibitem[2002]{uchiyama} Uchiyama, Y.; Takahashi, T.; Aharonian, F. A.;
 Mattox, J. R., 2002, ApJ, 571, 866

   \bibitem[]{} Wang, N. et al., 2001, MNRAS, 328, 855  

   \bibitem[]{} Watson, M. G. et al., 2009, A\&A, 493, 339

   \bibitem[2006]{weisskopf} Weisskopf, M. C. et al., 2006,
      ApJ, 652, 387

\end{thebibliography}
\end{document}